\documentclass[namedreferences]{solarphysics}

\usepackage[hyperref,optionalrh,showbiblabels]{spr-sola-addons} 
\usepackage{graphicx}        
\usepackage{color}           
\usepackage{breakurl}        
\usepackage{natbib}

\renewcommand{\vec}[1]{{\mathbfit #1}}

\newcommand{\aap}{    {\it Astron. Astrophys.}}

\newcommand{\apj}{    {\it Astrophys. J.}}
\newcommand{\apjl}{   {\it Astrophys. J. Lett.}}

\newcommand{\grl}{    {\it Geophys. Res. Lett.}}

\newcommand{\jgr}{    {\it J. Geophys. Res.}}

\newcommand{\ssr}{    {\it Space Sci. Rev.}} 
\chardef\us=`\_

\begin{document}

\begin{article}
\begin{opening}

\title{Nonlinear Evolution of Ion Kinetic Instabilities in the Solar Wind}

\author[addressref={aff1,aff2,aff3},corref,email={ofman@cua.edu}]{\inits{L.O.}\fnm{L. Ofman}}

\address[id=aff1]{Department of Physics, Catholic University of America, Washington, DC 20064, USA}
\address[id=aff2]{Code 671, NASA Goddard Space Flight Center, Greenbelt, MD 20771, USA}
\address[id=aff3]{Visiting, Tel Aviv University, Tel Aviv, Israel}

\runningauthor{Ofman}
\runningtitle{Nonlinear Evolution of Ion Kinetic Instabilities in the Solar Wind}

\begin{abstract}
In-situ observations of the solar wind (SW) plasma from 0.29 to 1AU show that the protons and
$\alpha$ particles are often non-Maxwellian, with evidence of kinetic instabilities, temperature anisotropies,
 differential ion streaming, and associated magnetic fluctuations spectra. The kinetic instabilities in the
SW multi-ion plasma can lead to preferential heating of $\alpha$ particles and the dissipation of magnetic fluctuation energy, affecting the kinetic and global properties of the SW. Using for the first time a three-dimensional hybrid model,  where ions are modeled as particle using the Particle-In-Cell (PIC) method and electrons are treated as fluid, we study the onset, nonlinear evolution and dissipation of ion kinetic instabilities. The Alfv\'{e}n/ion-cyclotron, and the ion drift instabilities are modeled in the region close to the Sun ($\sim10R_s$). Solar wind expansion is incorporated in the model. The model produces self-consistent non-Maxwellian velocity
distribution functions (VDFs) of unstable ion populations, the associated temperature anisotropies, and wave spectra for several typical SW instability cases in the nonlinear growth and saturation stage of the instabilities.  The 3D hybrid modeling of the multi-ion SW plasma could be used to study the SW acceleration region close to the Sun that will be explored by the Parker Solar Probe mission.

\end{abstract}
\keywords{Solar Wind, Theory: Numerical Modeling; Instabilities; Waves, Plasma}
\end{opening}

\section{Introduction}
  \label{int:sec}
 
In-situ observations of the solar wind (SW) plasma at distances $>$0.29AU by {\it Helios}, {\it Ulysses}, ACE, and {\it Wind} spacecraft show that the $\alpha$ particle population is usually hotter than that of the proton population, and flows faster by an Alfv\'{e}n speed in the fast SW streams. The VDFs of protons and ions exhibit non-Maxwellian features, such as temperature anisotropy with respect to the background magnetic field, beams (with stronger departures from Maxwellian for the $\alpha$ particles), and differential ion streaming \citep{Mar82}. These effects are stronger in the fast SW streams (compared to slow SW), increasing in magnitude closer to the Sun. Kinetic instabilities, such as ion-cyclotron, mirror and firehose, play an important role in shaping the SW plasma properties, as evident from observations at 1AU with WIND \citep{Bal09,Mar12} and other spacecraft data.

 Observed break points in the magnetic fluctuations power spectra indicate the inertial and kinetic dissipation ranges of magnetic field fluctuations \citep[see the review,][]{BC13}, with
the break point of the dissipation range aligned with the proton cyclotron frequencies at the various heliocentric distances \citep[e.g.,][]{Bou12,Tel15}.  Direct evidence of electromagnetic ion cyclotron (EMIC) waves near the proton cyclotron frequency was found in the SW by STEREO at 1AU \citep[e.g.,][]{Jia09,Jia14} and from {\it Helios} and MESSENGER at 0.3 AU \citep{Jia10}. Preferential acceleration and heating of the $\alpha$ particle populations was associated with wave activity in SW observations  at 1AU \citep{Kas08,Kas13, Bou11a,Bou11b,Bou13,Mar11,Mar12,BG14,Bor16,Dur19}.

Recently launched NASA's {\it Parker Solar Probe} (PSP) mission, is going to provide detailed in-situ observations as close as 10$R_s$, aimed at studying the SW acceleration and heating processes \citep{Fox16}. In anticipation of the PSP SW ions observations \citep{Kas16} we developed a 3D hybrid model of SW plasma with protons and $\alpha$ particles and study the onset, evolution, and dissipation of ion driven kinetic instabilities, using our 3D hybrid kinetic, expanding computational box model (see model details below).

Recently, ion kinetic instabilities in  SW electron-proton-alpha ($e-p-\alpha$) plasma were studied extensively with 2.5D hybrid models, including effects such as SW expansion and background inhomogeneity, and turbulence   \citep[e.g.,][]{Ofm14,OOV15,MOV15,Ofm17,MP18}. The advantage of the hybrid models over other kinetic modeling methods, such as full PIC and Vlasov solvers is that they can model the nonlinear evolution of the ion kinetic instabilities for parameter ranges, physical size, and duration not accessible by other methods, relaxing the approximations on mass ratio and speed of light used in PIC methods (the trade-off is the limitation of hybrid modeling to plasmas where electron kinetic evolution or instabilities could be neglected or are uncoupled from the ions), and the severe resolution limitations in velocity space of multi-dimensional Vlasov codes \citep[e.g.,][]{Per14}. Three-dimensional hybrid codes were developed for the study of SW $e-p$ plasma turbulence with more general (than 2.5D) description of the wave-particle interactions \citep[e.g.,][]{Vas14,Vas15,Fra18}. Here, for the first time, we use 3D hybrid model of $e-p-\alpha$ SW plasma to demonstrate the anisotropic ion  heating as a result of ion drift and temperature anisotropy instabilities, and the associated high frequency Alfv\'{e}n/ion-cyclotron wave spectra  in the SW plasma using  parameters relevant to the heliocentric distances that will be sampled for the first time by the recently launched PSP mission. The paper is organized as follows: in Section~\ref{hyb:sec} we describe the details of the hybrid model, in Section~\ref{num:sec} we present the numerical results, and the summary and conclusions are in Section~\ref{con:sec}.


\section{Hybrid Model}
\label{hyb:sec}
We use the 3D hybrid model where the protons and $\alpha$ particles are described kinetically as particles using the Particle-In-Cell (PIC) method,  while electrons are treated as neutralizing massless background fluid. The parallelized 3D hybrid code is an extension of the 1.5D hybrid code initially developed by \citet{WO93}, and parallelized 2.5D hybrid models developed to study multi-ion SW plasma \citep[e.g.,][]{OV07,Ofm14,Ofm17}. The equations of motion  (with the usual notations for the variables) are solved for each particle of all ion species ($s$) subject to the Lorentz force for 3D position $\mbox{\bf x}$ and velocity $\mbox{\bf v}$ vectors:
\begin{eqnarray}
&&\frac{d\vec{x}_s }{dt} =\vec{v}_s\\
&&m_s \frac{d\vec{v}_s }{dt} =Z_se\left(\vec{E}+\frac{\vec{v}_s\times \vec{B}}{c} \right),
\label{motion:eq}\end{eqnarray} where $m_s$ is the particles mass, $Z_s$ is the charge number, $e$
is the electron charge, $c$ is the speed of light, $\vec{E}$ is the electric field, and $\vec{B}$ is the magnetic field. The electron momentum equation is solved in 3D by using negligible electron
inertia leading to the generalized Ohm's law
\begin{eqnarray}
&&en_{e} \left(\vec{E}+\frac{\vec{v}_{e} \times \vec{B}}{c} \right)+\vec{\nabla }p_{e}=0,
\label{ohm:eq}\end{eqnarray} where the scalar electron pressure equation of state $p_e =k_Bn_eT_e$ is used for closure, where $T_e$ is the electron temperature, and $n_e$ is the electron number density. For stability a resistive term $en_e\eta \vec{J}$ is added to the r.h.s. of Equation~\ref{ohm:eq}, where $\eta$ is an empirical small resistivity coefficient, and $\vec{J}$ is the local current density. The above equations are supplemented by the quasi-neutrality condition
$n_e =n_p+2n_\alpha$, where $n_p$ is the proton population and $n_\alpha$ is the $\alpha$ particle population  number density, respectively. Maxwell's equations $\nabla \times \vec{B}=\frac{4\pi }{c} \vec{J},$ and $\nabla \times \vec{E}=-\frac{1}{c} \frac{\partial \vec{B}}{\partial t} $ are solved on the 3D spatial grid, with grid sizes of up to $128^3$ and 64 to 128 particles per cell (ppc) in the present study. The proton and $\alpha$ particle equations of motions are advanced in time as the
particle motions respond to the updated Lorentz force at each time step. The particle and field equations are integrated in time
using the Rational Runge-Kutta (RRK) method \citep{Wam78}
whereas the spatial derivatives are calculated by pseudospectral
FFT method in 3D, and periodic boundary conditions are applied.  The equations of motions of the ions are solved using Cartesian coordinates with the background uniform magnetic field $B_0$ direction defined along ${\rm \bf x}$. The spatial coordinates are normalized using the proton inertia length $\delta=c/\omega_{pp}$, where $\omega_{pp}=(4\pi n_pe^2/m_p)^{1/2}$ is the proton plasma frequency,  the time is in units of the inverse proton cyclotron frequency, $\Omega_p^{-1}$, where $\Omega_p=\frac{eB_0}{m_pc}$, and the velocities are in units of the Alfv\'{e}n speed, $V_A$.  In the present study the spatial resolution is up to $0.75\delta$, and the time step is on the order of $0.01\Omega_p^{-1}$. The hybrid modeling 
method has been tested and used successfully in many previous studies, and details of the normalization were published \citep[see, e.g.,][]{WO93,OV07,Ofm10a,Ofm14,Ofm17}. 

\subsection{Expanding Box Model}
A solar wind plasma parcel expands naturally as it travels away from the Sun into the heliosphere at the SW speed, $U_0$. In order to model the expansion of the SW, we use the Expanding Box model (EBM), developed  initially by \citet{GV96} for an MHD fluid code and later adapted by \citet{LVG01} in their 1.5D hybrid code, \citep[see, also][]{Hel03,Hel05}. We have  implemented the EBM in our 2.5D hybrid code \citep{OVM11}, and recently applied for the first time to our 3D hybrid code. Here, we summarize the equations given in \citet{OVM11}, extended to the 3D hybrid model. 

The heliocentric radial position, $R(t)$, of the SW plasma parcel as a function of time is approximated by 
\begin{eqnarray}
&&R(t)=R_0+U_0t,
\end{eqnarray} where a constant SW velocity $U_0$ is assumed (i.e., applicable beyond $\sim10R_s$), and $R_0$ is the reference radial distance. The dimensionless expansion factor $a(t)$ is  
\begin{eqnarray}
&&a(t)=R(t)/R_0=1+\frac{U_0}{R_0}t.
\end{eqnarray} The Galilean transformation $v'=v-U_0$ is used to transform to the SW plasma rest frame. Since the expansion rate of the SW is 'slow' the expansion parameter is small, $\epsilon=\frac{U_0}{R_0}t\ll 1$, where time is in units of $\Omega_p^{-1}$. Thus, the coordinates are transformed to the moving frame as follows:
\begin{eqnarray}
&&x=x'+R,\ y=ay',\ z=az'.
\end{eqnarray} While the $x$ coordinate (along the radial direction) undergoes Galilean transformation, the $y$ and $z$ coordinates undergo stretching, or inflation. Thus, the ion equation of motions in  the 3D hybrid model transforms as  
\begin{eqnarray}
&&\frac{d\mbox{\bf v}}{dt}=\frac{d\mbox{\bf v}'}{dt'}+\frac{\dot{a}}{a}\mbox{\bf P}\cdot\mbox{\bf v}',\ \mbox{\bf P}= \left( \begin{array}{ccc}
0 & 0 & 0\\
0 & 1 & 0 \\
0 & 0 & 1 \end{array} \right),
\end{eqnarray} where we note that for $\epsilon\ll1$, we can use the approximation $\frac{\dot{a}}{a}\sim\frac{U_0}{R_0}$, and where {\bf v}$'$ components are given by 
\begin{eqnarray}
&&\frac{dx'}{dt}=v_x',\ \frac{dy'}{dt}=\frac{1}{a}v_y',\ \frac{dy'}{dt}=\frac{1}{a}v_y'.
\end{eqnarray} 
For the magnetic field we have
\begin{eqnarray}
&&\frac{\partial \mbox{\bf B}'}{\partial t'}+\mbox{\bf B}'(\nabla'\cdot\mbox{\bf U}_i')-(\mbox{\bf B}'\cdot\nabla')\mbox{\bf U}_i'+(\mbox{\bf U}_i'\cdot\nabla')\mbox{\bf B}'\\
&&+\frac{c}{4\pi e n_e}\nabla'\times\left[(\nabla'\times\mbox{\bf B}')\times\mbox{\bf B}'\right]=-\frac{2U_0}{R_0}B_x'\hat{x}-\frac{U_0}{R_0}B_y'\hat{y}-\frac{U_0}{R_0}B_z'\hat{z},\nonumber
\end{eqnarray} where the transformed ion bulk velocity is $\mbox{\bf U}_i'=\mbox{\bf U}_i-\mbox{\bf U}_0$, and the electric field in the moving frame are 
\begin{eqnarray}
\ \ \mbox{\bf E}'=&-&\frac{1}{c}\mbox{\bf U}_i'\times\mbox{\bf B}'+\frac{1}{4\pi e n_e}(\nabla'\times\mbox{\bf B}')\times\mbox{\bf B}'-\frac{1}{en}\nabla'(n_ek_BT_e),
\end{eqnarray}
The above transformations are implemented in the 3D hybrid code, including the derivatives with respect to $x'$, $y'$ and $z'$, with the expansion parameter $\epsilon$. The typical expansion parameter, $\epsilon$ at $10R_s$ is $10^{-4}-10^ {-5}$, while setting $\epsilon=0$, recovers the non-expanding model. In order to evaluate the effects of expansion, we present preliminary results with a higher expansion rate, due to computational limitations of the  3D hybrid model, while more realistic (smaller) values of $\epsilon$ requiring longer runs left for a future study.

\section{Numerical Results}
\label{num:sec}
\begin{figure}[h]
\begin{center}
\includegraphics[width=8.5cm]{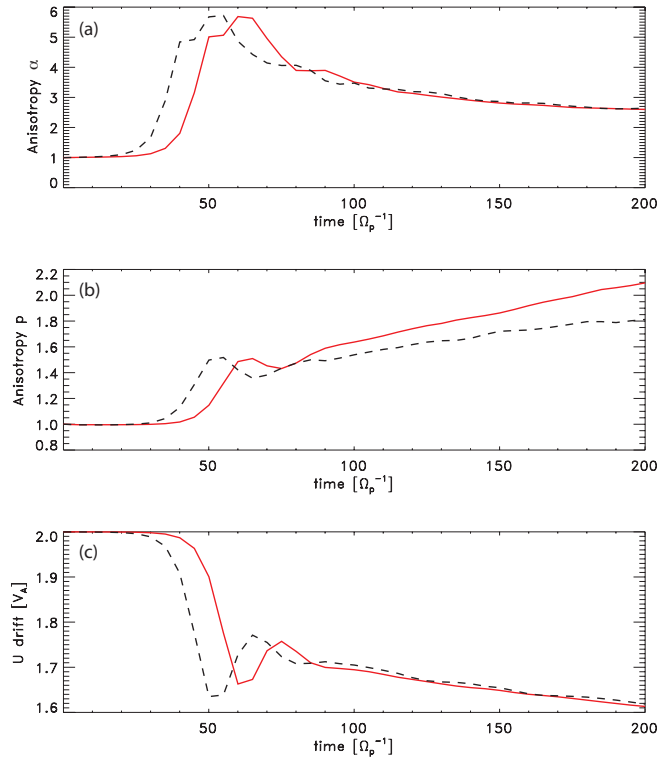}
\end{center}
\caption{The results of a test run that compares the temporal evolution of the $\alpha$-$p$ drift instability modeled with the 3D hybrid code ({\it red}, Case~1) and the 2.5D hybrid code ({\it black dashed line}). (a) The temperature anisotropy of the $\alpha$ population. (b) The temperature anisotropy of the protons. (c) The relative $\alpha$-$p$ fluid drift velocity calculated from the VDFs.}
\label{3Dvs2D:fig}
\end{figure}

\begin{figure}[ht]
\begin{center}
\includegraphics[width=7cm]{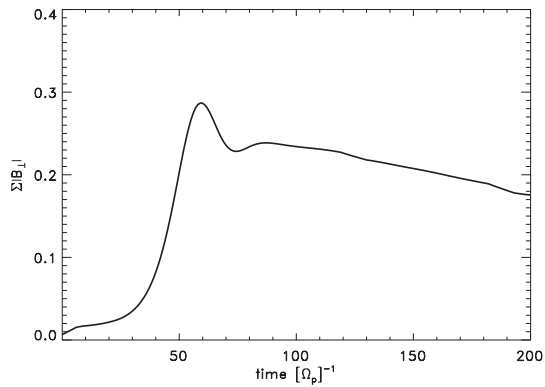}
\end{center}                 
\caption{The temporal evolution of the perpendicular magnetic fluctuations power $|B_\perp|^2$ produced in the 3D hybrid model of Case~1.}
\label{bperp_drift:fig}
\end{figure}

\begin{figure}[ht]
\begin{center}
\includegraphics[width=12cm]{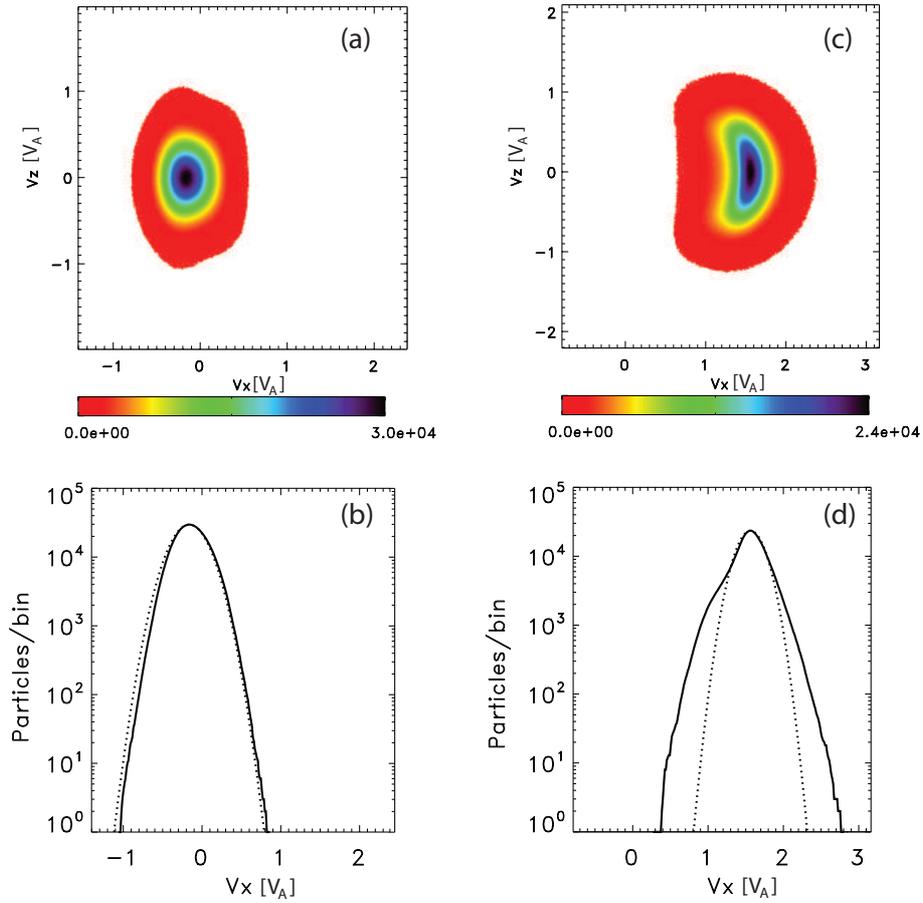}
\end{center}                 
\caption{The VDFs of the proton and $\alpha$ particle populations obtained from the 3D hybrid model for the drift instability shown in Figure~\ref{3Dvs2D:fig} (Case~1) at $t =90\Omega_p^{-1}$ for the non-expanding case. (a) $V_x-V_z$ of protons. (b) The cut through the VDF at $V_z=0$ showing the non-Maxwellian features in the $V_x$ VDF of protons ({\it solid}). The best-fit Maxwellian is shown ({\it dashes}). (c) The cut through the VDF at $V_z=0$ showing the non-Maxwellian features in the $V_x$ VDF of $\alpha$ particles ({\it solid}). The best-fit Maxwellian is shown ({\it dashes}).}
\label{VDF_drift:fig}
\end{figure}

\begin{figure}[h]
\begin{center}
\includegraphics[width=10cm]{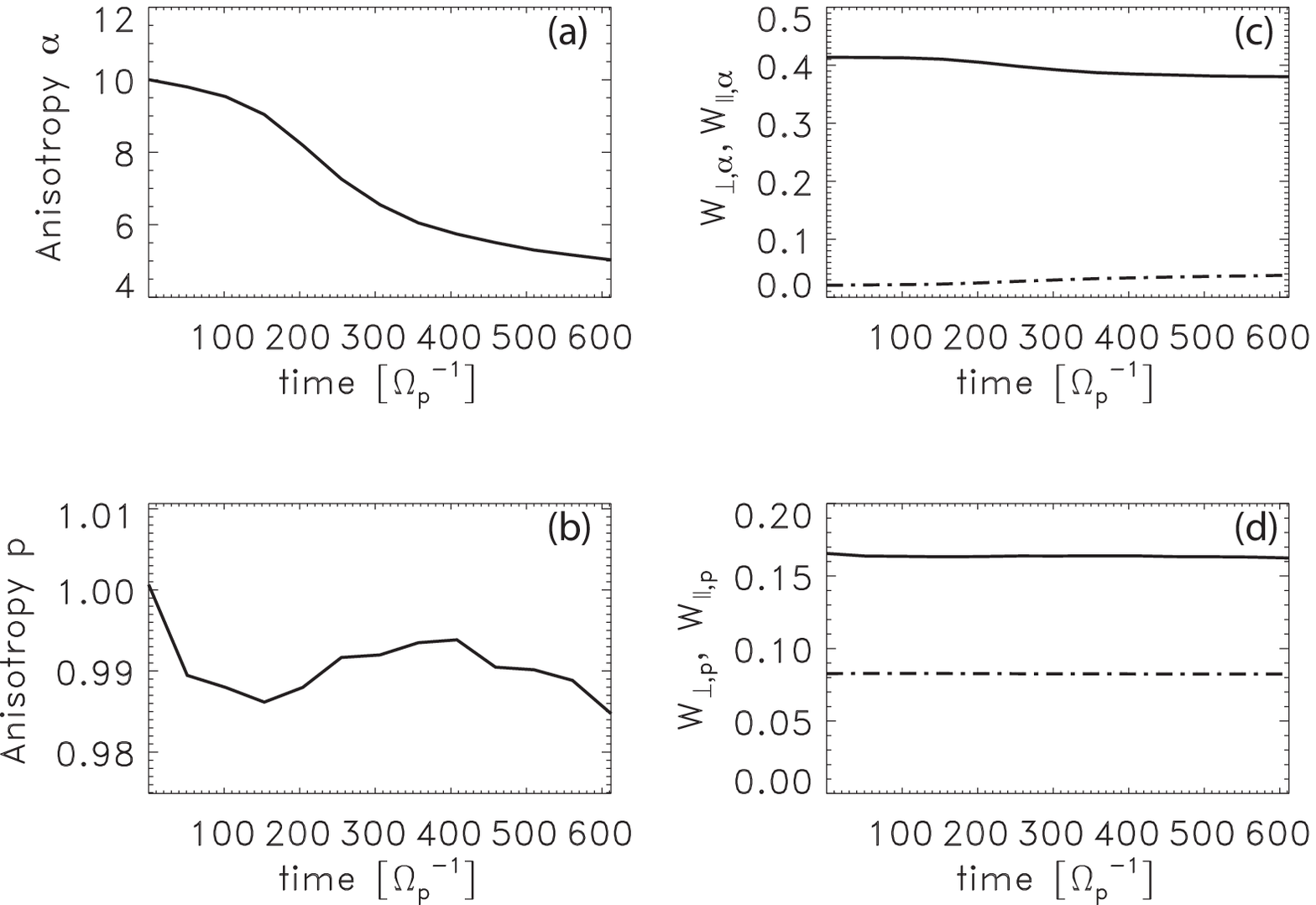}
\end{center}
\caption{The relaxation of the $\alpha$ particle population temperature-anisotropy-driven ion-cyclotron instability modeled with the 3D hybrid code (Case~2). The initial $T_{\perp,\alpha}/T_{\parallel,\alpha}=10$, while protons are isotropic with initial $\beta_{\parallel,\alpha}=\beta_{\parallel,p}=0.04$. (a) The temporal evolution of the $\alpha$ particle population temperature anisotropy. (b) The temporal evolution of the proton temperature anisotropy. (c) Parallel and perpendicular kinetic energies (in normalized units) of the $\alpha$ particles, $W_{\parallel,\alpha}$ ({\it dot-dashes}),  $W_{\perp,\alpha}$ ({\it solid}). (d) Parallel and perpendicular kinetic energies (in normalized units) of the protons, $W_{\parallel,p}$ ({\it dot-dashes}),  $W_{\perp,p}$ ({\it solid}).}
\label{Aniso10:fig}
\end{figure}

\begin{figure}[ht]
\begin{center}
\includegraphics[width=8cm]{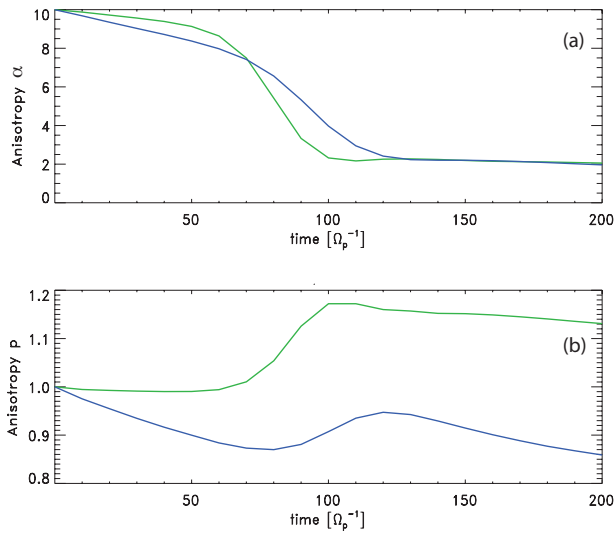}
\end{center}
\caption{The temporal evolution of the temperature anisotropy of (a) $\alpha$ particle population, and (b) protons driven by initial $\alpha$ population temperature anisotropy with initial $\beta_{\parallel,\alpha}=\beta_{\parallel,p}=1$ (note the different scale of the $y$-axis). The non-expanding solutions are shown ({\it green}, Case~3), and the effect of expansion with $\epsilon=10^{-3}$ are shown ({\it blue}, Case~4). }
\label{expansion:fig}
\end{figure}

\begin{figure}[ht]
\begin{center}
\includegraphics[width=12cm]{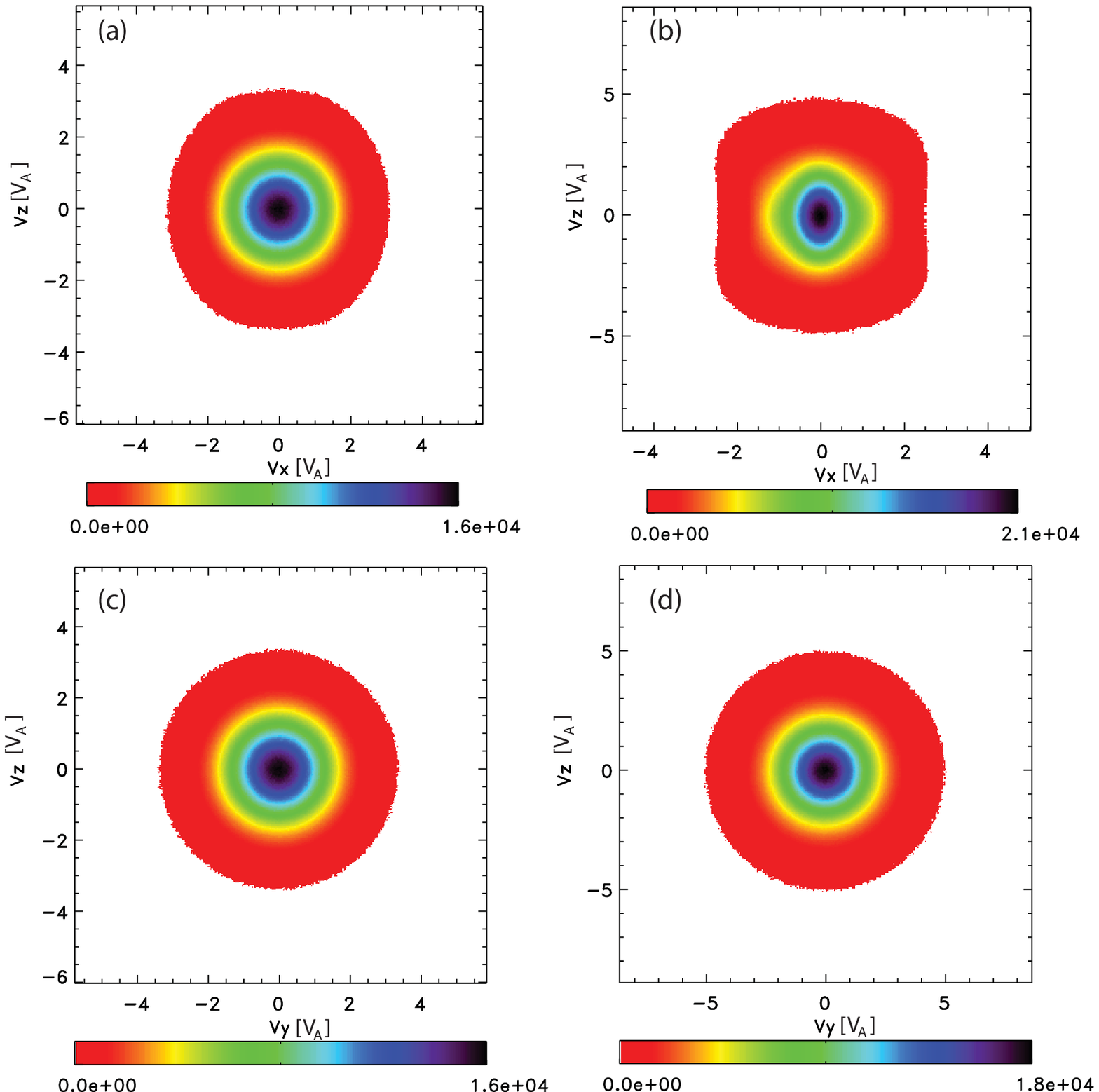}
\end{center}                 
\caption{The VDFs of the proton and $\alpha$ particle populations obtained from the 3D hybrid model for the ion-cyclotron instability shown in Figure~\ref{expansion:fig} at $t =200\Omega_p^{-1}$ for the non-expanding model  (Case~3). (a) $V_x-V_z$ of protons. (b) $V_x-V_z$ of $\alpha$ particles. (c) $V_y-V_z$ of protons. (b) $V_y-V_z$ of $\alpha$ particles. }
\label{VDF_IC:fig}
\end{figure}

\begin{figure}[ht]
\begin{center}
\includegraphics[width=10cm]{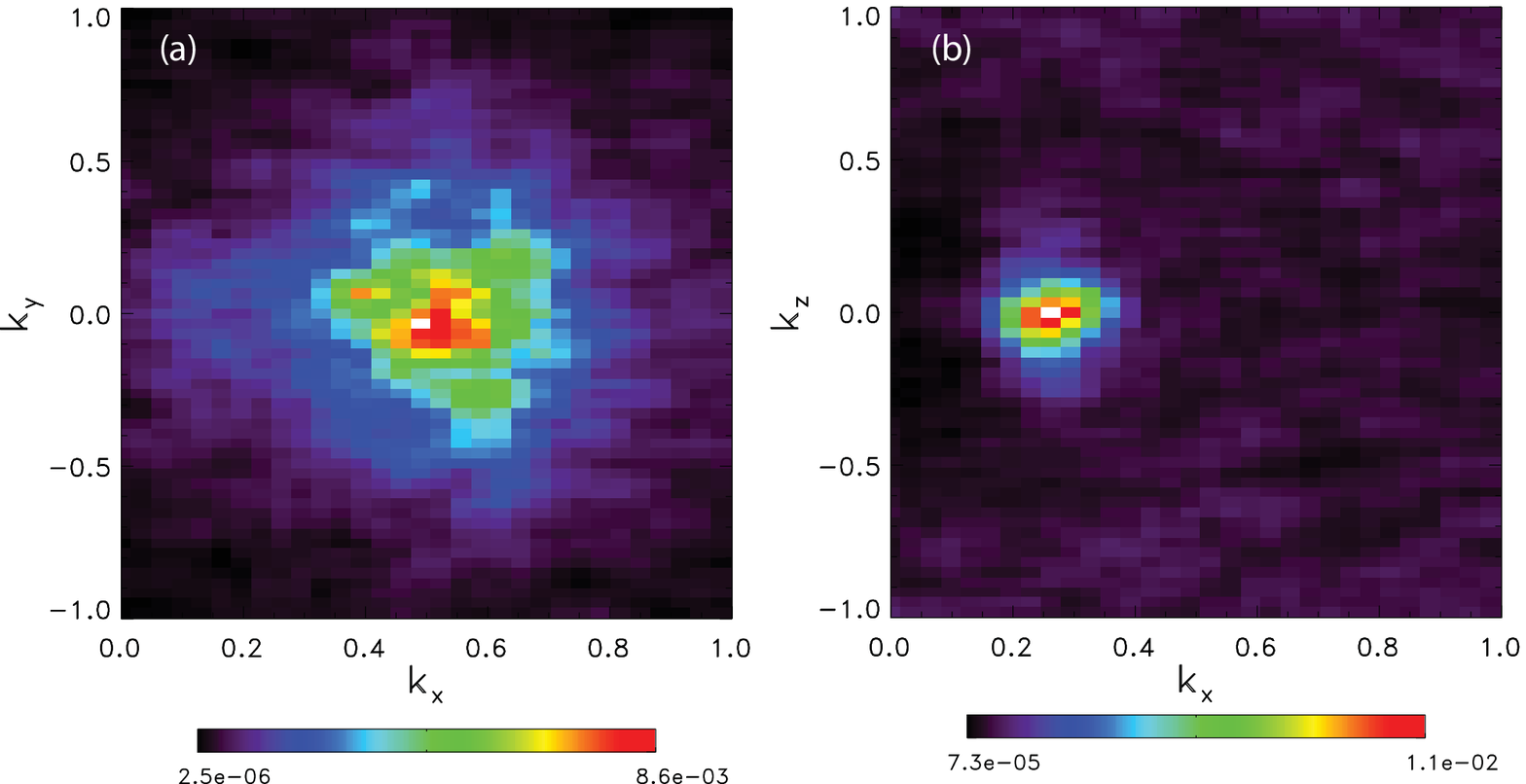}
\end{center}                 
\caption{The $k$-space power spectra of the magnetic fluctuations due to the instabilities modeled here with the 3D hybrid code. (a) The $\alpha$-proton drift instability with initial drift $V_d=2V_A$ (Case~1)  at $t=200\Omega_p^{-1}$. (b) The ion-cyclotron instability with initial $\beta_\parallel=1$ (Case~3) at $t=119\Omega_p^{-1}$. The wavenumber $k$ is normalized in units of $\delta^{-1}$.}
\label{k_spect:fig}
\end{figure}

In Figures~\ref{3Dvs2D:fig}-\ref{k_spect:fig} we present the results of the 3D hybrid modeling for the cases with parameters summarized in Table~1. The parameters of the instabilities in this study exceed the linear stability threshold in all cases as determined form solutions of Vlasov's linear dispersion relation \citep[see, e.g.,][]{DO75,Gar93,Gar01,Gar03,XOV04,OV07}. We have used 5\% $\alpha$ particles number density (in terms of $n_e$). In Figure~\ref{3Dvs2D:fig} we compare the evolution of the ion drift instability (Case~1), modeled with both, 2.5D and 3D hybrid codes with identical parameters. It is evident that the relaxation of the drift instability is rapid in terms of proton gyro-periods, and the drift velocity relaxes from 2 to $\sim$1.6 in 200$\Omega_p^{-1}$. At the same time, the ion populations are heated in the perpendicular direction and becomes anisotropic with maximal $T_{\perp,p}/T_{\parallel,p}\approx2.1$, and $T_{\perp,\alpha}/T_{\parallel,\alpha}\approx5.6$. The 2.5D and 3D hybrid models produce similar results, with the initial growth somewhat slower an higher final proton anisotropy in the 3D hybrid model. The final states of $\alpha$ population temperature anisotropy and drift velocity are close in both models. The similarity in the evolution is expected, since most of the power in the modes are  in the parallel (to the magnetic field) direction. These results provide additional validation of our newly develop 3D hybrid model.

In Figure~\ref{VDF_drift:fig} we show the  velocity distribution functions (VDFs) of the protons and $\alpha$ particle  populations for Case~1 at $t=90\Omega_p^{-1}$. The anisotropy and the non-Maxwellian features are evident in the $V_x-V_z$ velocity space of both ion species VDFs. The larger (than proton) anisotropy is seen in the $\alpha$ population VDF. In the lower panel of Figure~\ref{VDF_drift:fig} we show cuts of the VDFs at $V_z=0$, obtained by integrating the number of particles in velocity bins (i.e., $ \Delta V_x\ll1$) for each species, and the best-fit Maxwellians. It is evident that protons are nearly Maxwellian in the parallel direction, while $\alpha$ particle population has a Maxwellian core with a non-thermal tail, likely produced by the nonlinearity of the drift instability, i.e., wave-particle interactions and pitch-angle scattering populating the VDF tails. The 3D hybrid modeling results are similar to the previous 2.5D hybrid modeling results of this instability \citep[see, e.g.,][]{Ofm14}.

The temporal evolution of the perpendicular magnetic fluctuations power, $|B_\perp|^2$, are shown in Figure~\ref{bperp_drift:fig}. The perpendicular fluctuations are primarily due to the wave power as demonstrated in previous studies (for example, see the dispersion relation in Figure~12 in 
\citet{Ofm17} obtained from 2.5D hybrid model). The initial rapid growth of the  magnetic power accompanies the initial growth of the drift instability evident in Figure~\ref{3Dvs2D:fig}. The peak perpendicular magnetic power is reached at $t\approx55\Omega_p^{-1}$, followed by nonlinear saturation and gradual dissipation. This is consistent with the perpendicular heating of the ions, as evident in the evolution of their anisotropies in  Figure~\ref{3Dvs2D:fig}, suggesting the Alfv\'{e}n/cyclotron wave absorption as the heating process.

The evolution of the temperature anisotropies, and parallel and perpendicular kinetic energies of protons and $\alpha$ particle populations for Case~2 are shown in Figure~\ref{Aniso10:fig}. The ion-cyclotron instability is driven by the initial temperature anisotropy of the $\alpha$ particles, $T_{\perp,\alpha}/T_{\parallel,\alpha}$=10, while protons are initially isotropic. It is evident that the temperature anisotropy of the $\alpha$ population is gradually decreasing with time, while protons remain nearly isotropic (note the small $y$-axis range of the proton anisotropy plot). 

The decrease of $\alpha$ temperature anisotropy is due to the decrease of their perpendicular kinetic energy, accompanied by an increase of the parallel energy due to velocity phase space diffusion. At the same time, the proton parallel and perpendicular kinetic energies remain nearly constant.  At $t=600\Omega_p^{-1}$ the  $\alpha$ particle population temperature anisotropy has decreased by a factor of two to $\sim5$, approaching gradually the equilibrium state. Due to computational limitation, the run was not continued for a longer duration.

The temporal evolution of the temperature anisotropy of proton and $\alpha$ particle populations for Cases~3 and 4 (Table~1) are shown in Figure~\ref{expansion:fig}. The initial $T_{\perp,\alpha}/T_{\parallel,\alpha}=10$ and initial $\beta_{\parallel,\alpha}=\beta_{\parallel,p}=\beta_e=1$ (where we use $n_e$ in the definition of $\beta$ of all species). As expected from linear theory, the relaxation of the ion-cyclotron instability driven by the $\alpha$ temperature anisotropy is faster in this case compared to the low-$\beta_\parallel$ case shown in Figure~\ref{Aniso10:fig}. The relaxation  of the $\alpha$ temperature anisotropy occurs in $\sim100\Omega_p^{-1}$ and an asymptotic state with $T_{\perp,\alpha}/T_{\parallel,\alpha}\approx 2$ is reached and remains to the end of the run. The protons are initially isotropic and undergo perpendicular heating due to the absorptions of the wave-spectrum produced by the relaxation of the $\alpha$ population. The maximal proton anisotropy is $\sim1.17$ for the non-expanding case, relaxing towards isotropic state. When gradual expansion is introduced in the model ($\epsilon=10^{-3}$), it is evident that the relaxation of the temperature anisotropy is more rapid than in the non-expanding case, with further perpendicular cooling of the protons. This is expected due to the `stretching' of the perpendicular coordinates as the parcel of plasma expands in the heliosphere, and consistent with previous 2.5D hybrid EBM modeling results \citep[e.g.,][]{OVM11,Ofm14}.

The VDFs of the proton and $\alpha$ particle populations for the ion-cyclotron instability (Case~3) are shown in Figure~\ref{VDF_IC:fig} at the end of the evolution at $t=200\Omega_p^{-1}$ near the asymptotic quasi-steady state. The anisotropy and the non-Maxwellian features are evident in both proton and $\alpha$ particle VDFs in the $V_x-V_z$ phase space, with the larger anisotropy of $\alpha$ particles. The perpendicular velocity phase-space plane is nearly Maxwellian in both species, as evident in the circular shapes of the VDFs. This property is due to the dominance of the parallel propagating modes, with little effect on the perpendicular direction (see, Figure~\ref{k_spect:fig}, below). The results are consistent with previous 2.5D modeling studies of this instability \citep[e.g.,][]{OVM11}.

The 2D power spectra of the magnetic fluctuations for the instabilities (Case~1 and Case~3) obtained from the 3D hybrid model are shown in Figure~\ref{k_spect:fig}. It is evident that for both cases the peak power of the magnetic fluctuations is located near the $k_x$ axis, while the power at $|k_\perp|>0$ is significantly diminished. There is evidence for somewhat larger power in the oblique modes for the drift instability (Case~1), compared to the ion-cyclotron instability (Case~3). This result is consistent with the dominant growth of the parallel propagating modes expected from solutions of Vlasov's linear dispersion relation \citep[e.g.,][]{OV07}. We have found that as the instabilities evolve and dissipate, the power at higher $k$ is dissipated, and the peak power in $k$-space is moving towards the origin in $k$-space (i.e., longer wavelengths). 

\begin{table*}
\caption{The parameters of the initial states for the 3D hybrid modeling cases in the present study. In all cases the proton population is initially anisotropic, $\beta_\parallel=\beta_{\parallel,p}=\beta_{\parallel,\alpha}=\beta_e$, and $n_\alpha=0.05n_e$.}
\begin{tabular}{clcccc}
\hline
	Case        &  Instability type  &$\beta_\parallel$  &  $T_{\perp,\alpha}/T_{\parallel,\alpha}$  & $V_d [V_A]$ & $\epsilon$ \\ \hline
	           1      & Drift                  &   0.04          &      1   & 2 & 0 \\
          2               & Ion-Cyclotron   & 0.04            &      10 &  0 & 0\\
           3              & Ion-Cyclotron   & 1             &    10  &0 & 0 \\
          4               & Ion-Cyclotron    & 1            &    10  &  0  &$ 10^{-3}$
\end{tabular}
\label{model_par}
\end{table*}

\section{Summary and Conclusions} 
      \label{con:sec} 
 
Motivated by past SW ion observations at 1AU and the inner heliosphere and by anticipated data from the PSP mission close to the Sun, we investigate kinetic instabilities in $e-p-\alpha$ SW plasma. The instabilities are initialized with temperature anisotropy ($T_\perp>T_\parallel$) of $\alpha$ particle populations, and an initial relative proton-$\alpha$ super-Alfv\'{e}nic drift modeled for the first time with full 3D hybrid simulations and EBM implementation. We extend previous 2.5D hybrid modeling studies to more realistic 3D model and find general agreement with previous 2.5D result, with 3D effects affecting properties, such as the growth or relaxation rate of the instabilities in the nonlinear stage. 

The instability leads to rapid relaxation of the super-Alfv\'{e}nic drift, producing EMIC waves and associated perpendicular magnetic fluctuations power. The perpendicular velocity distributions become non-Maxwellian and $\alpha$ particle population is heated with strong perpendicular temperature anisotropy similar to previous 2.5D studies and spacecraft observations.
We find that the SW protons are heated significantly in the perpendicular direction by the ion drift instability and the associated power of kinetic waves. We calculate the $k$-space power spectra of the waves and find that they are dominated by parallel propagating waves, with a small oblique component.
 
We demonstrate the various forms of self-consistent non-Maxwellian VDFs of protons and $\alpha$ particle populations produced by the modeled instabilities and  wave-particle interactions with the associated wave spectra  in the nonlinear and relaxed stages of the evolution. The  results of the study are relevant to the understanding of the kinetic wave-particle processes that likely take place in the  acceleration region of the SW multi-ion plasma close to the Sun ($\sim10R_s$), the target investigation region of the recently launched NASA's PSP mission, where the study of heating and acceleration of proton and $\alpha$ populations is one of the major goals.
 
\begin{acks}
 The author acknowledges support by NASA cooperative agreement NNG11PL10A 670.154 to Catholic University of America. Resources supporting this work were provided by the NASA High-End Computing (HEC) Program through the NASA Advanced Supercomputing (NAS) Division at Ames Research Center.  
\end{acks}\\

\noindent{\bf Disclosure of Potential Conflicts of Interest} The author declares he has no conflicts of interest.




\end{article} 

\end{document}